\documentclass{sig-alternate}

\usepackage{url}
\usepackage{mathptmx}

\begin{document}
\title{Will 5G See its Blind Side? \\Evolving 5G for Universal Internet Access}

\numberofauthors{3} 
\author{
\alignauthor
Oluwakayode Onireti, Muhammad Ali Imran\\
 \affaddr{Institute for Communications System (ICS), University of Surrey, United Kingdom}\\
\email{o.s.onireti@surrey.ac.uk; m.imran@surrey.ac.uk}
\alignauthor Junaid Qadir\\
 \affaddr{Information Technology University (ITU)-Punjab, Lahore, Pakistan}\\\email{junaid.qadir@itu.edu.pk} 
\alignauthor Arjuna Sathiaseelan\\
 \affaddr{Computer Laboratory, University of Cambridge, United Kingdom}\\
\email{arjuna.sathiaseelan@cl.cam.ac.uk} 
}
 
\maketitle
\begin{abstract}
Internet has shown itself to be a catalyst for economic growth and social equity but its potency is thwarted by the fact that the Internet is off limits for the vast majority of human beings. Mobile phones---the fastest growing technology in the world that now reaches around 80\% of humanity---can enable universal Internet access if it can resolve coverage problems that have historically plagued previous cellular architectures (2G, 3G, and 4G). These conventional architectures have not been able to sustain universal service provisioning since these architectures depend on having enough users per cell for their economic viability and thus are not well suited to rural areas (which are by definition sparsely populated). The new generation of mobile cellular technology (5G), currently in a formative phase and expected to be finalized around 2020, is aimed at orders of magnitude performance enhancement. 5G offers a clean slate to network designers and can be molded into an architecture also amenable to universal Internet provisioning. Keeping in mind the great social benefits of democratizing Internet and connectivity, we believe that the time is ripe for emphasizing universal Internet provisioning as an important goal on the 5G research agenda. In this paper, we investigate the opportunities and challenges in utilizing 5G for global access to the Internet for all (GAIA). We have also identified the major technical issues involved in a 5G-based GAIA solution and have set up a future research agenda by defining open research problems.

\end{abstract}

\sloppy
\section{Introduction}

It is well known that Internet can facilitate economic growth; enable access to information; and create human convenience through digital networked services. Unfortunately, the Internet is inaccessible to a billions of human beings (it is estimated that 4 billion people---56\% of the world's population---are still without the Internet access \cite{A4A1}). There is a consensus emerging worldwide that the Internet is a basic human right and that no one should be denied access to the digital dividend that the Internet affords. This motivates the agenda for ensuring global access to the Internet for all (GAIA). The importance of GAIA can be gauged from the fact that the goal of providing \textit{universal and affordable Internet access} for everyone, everywhere by 2020 has been enshrined by the United Nations (UN) in 2015 as as one of the 17 sustainable development goals (SDG)  \cite{A4A1}.
 
The ubiquitous mobile cellular technology---which reaches 80\% of the human beings worldwide---can enable universal Internet access if it can resolve affordability and coverage problems that have precluded universal service provisioning using traditional cellular architectures (2G, 3G, and 4G). The 5th generation (5G) of mobile cellular networking, currently in a formative phase and expected to materialize by 2020, will usher in a new era of high-speed scalable mobile services beyond the current 4G standards. While the exact form of 5G has not yet emerged, 5G will definitely be a paradigm shift since the massive increase in data traffic requirements by 2020 will require a rethink of all aspects of the cellular architecture \cite{COST2013,Andrews2014}. The goals envisioned for 5G include a 1000$\times$ increase in capacity; an edge rate of 100 Mbps; a peak data rate in the range of tens of Gbps; round trip latency of 1 ms; and better energy efficiency in terms of 100$\times$ increase in the bits per joule. 

\subsection{Motivating a 5G GAIA Agenda}

\vspace{1mm}
``\textit{The digital divide is widening and is arguably of much larger concern than a local tenfold capacity increase in downtown Manhattan or in the streets of Tokyo}.'' \cite{erikssonanyone}
\vspace{1mm}

Traditionally cellular systems have had an overbearing focus on the peak data rate and have underempahsized universal coverage. Amidst the highly ambitious 5G 1000$\times$ capacity and rate targets defined for 5G, the coverage goals have been modest. For instance, the recent 2015 white paper from Next Generation Mobile Networks Alliance (NGMN) does not push for universal coverage, instead proposing a target of service availability 95\% of the time at 95\% of the locations \cite{NGMN2015}. This lack of attention to universal coverage is frustrating for critical applications such as emergency services that cannot tolerate failing connectivity, and in this vein, 5G will be unable to bridge the vast digital divide that has resulted between those who have and those who do not have access to information and communications technology (ICT). 

\begin{figure}
\centering
\includegraphics[width=.45\textwidth]{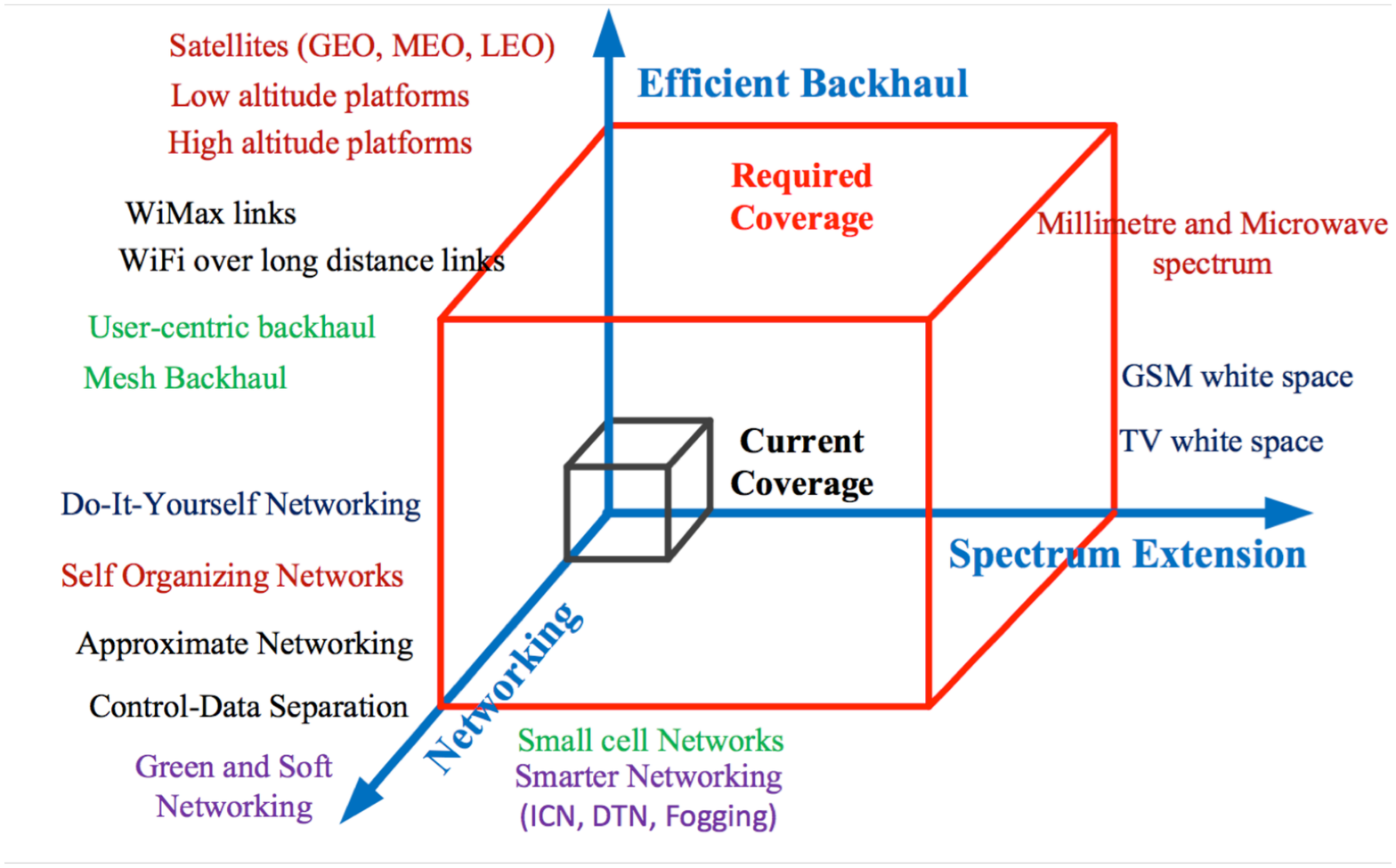}
\caption{Roadmap to Universal Internet Access}
\label{Fig_Fig_Road}
\end{figure}

In worldwide terms, the penetration of Internet technology is dismal (with only 1 in 5 people being connected to the Internet, according to 2015 statistics \cite{WDR2016}). Unfortunately, the bulk of the disconnected are those who are desperately poor---and the fact that they are not able to connect to the Internet creates a vicious cycle that keeps them in poverty since the digital divide deprives them of an equal growth opportunity. With the 100$\times$ speed increase envisioned in 5G, the digital divide will continue to widen---since technology amplifies and strengthens differences (the ``rich get richer'' phenomena also known as the \textit{Matthew's Law}). This widening digital divide should enliven the 5G community and catalyze an urgent development of a 5G GAIA research agenda, which shall help ensure that the benefit of 5G is accessible to everyone everywhere. To address the universal coverage gap in the 5G effort, we need to separately address solutions related to \textit{backhaul}, \textit{spectrum extension} and \textit{networking}, as illustrated in Figure \ref{Fig_Fig_Road}. 


%
%

\subsection{Challenges in 5G Based GAIA}

Since mobile operators are incentivized by profit making, their services are generally limited to dense areas where their revenue will compensate their capital expenditure (CAPEX) and their operating expenditure (OPEX). With this predicament, mobile operators may not be willing to provide coverage to the rural areas as they are largely sparsely populated and often lack the basic telephony infrastructure. The economic problems with GAIA is not only in provisioning of the networking service. It is estimated that around 80--90\% of the world's population lives within the range of 2G/3G mobile signal, with 3G covering approximately 69\% of the world's population; the reachability of mobile broadband is only around 10\% of the world's population  \cite{WDR2016}. The fact that a much lesser percentage of population accesses the Internet (despite the availability of service) highlights \textit{an affordability gap} which underscores the need for affordable service (and helps emphasize the fact that just because the network is there does not mean that people will necessarily be able to afford it). 

\subsection{Contribution of the paper}

The main contribution of this paper is to present how 5G can provision GAIA. We review a range of possible solutions that can be implemented to improve rural coverage. This includes  discussions on new technologies, such as user-centric solutions, context aware tradeoffs and control-data separation architecture solutions, and more radical exploitation of existing solution, such as satellites, aerial platforms and community wireless networking solutions. 

\subsection{Organization of the paper }

The rest of the paper is organized as follows. In Section 2, we present the building blocks of a possible 5G-based GAIA architecture. In Section 3, we provide a discussion on 5G GAIA issues---including an evaluation of various promising 5G technologies (such as ultra-densification, mmWave, and massive MIMO) for GAIA; a discussion on using context-appropriate tradeoffs for providing approximate networking; and the importance of energy efficiency for 5G-based GAIA. We finally conclude the paper in Section 4.

\section{5G Based GAIA Architecture}

\begin{figure}
\centering
\includegraphics[width=.45\textwidth]{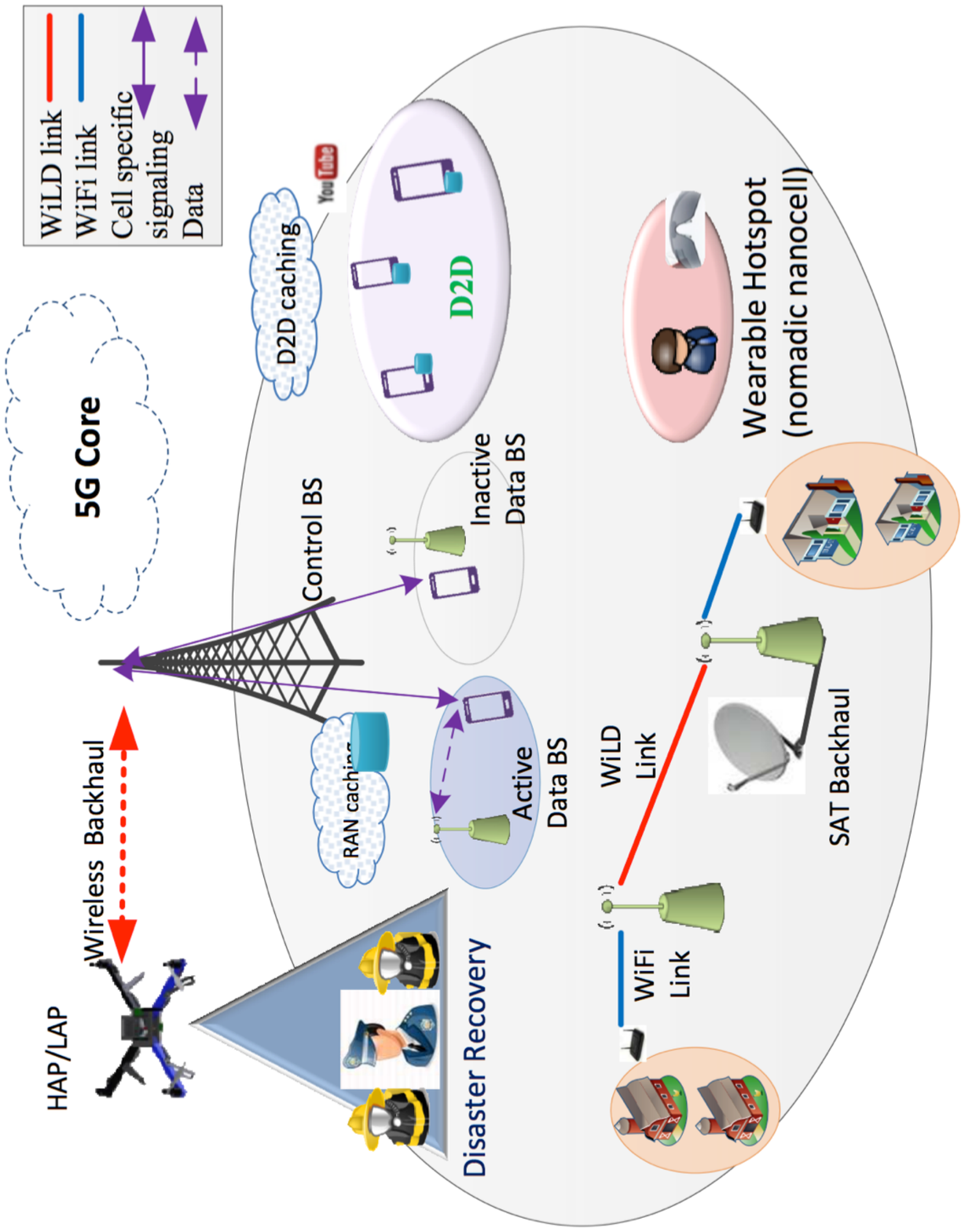}
\caption{The diversity of technology, applications, and techniques in 5G GAIA scenarios.}
\label{Fig_Fig_Sce}
\end{figure}

Far from being based on a single technology, the Internet and the mobile network of tomorrow will be an all-inclusive architecture that will subsume a number of different architectures and technologies. Already the designers of 5G technology are envisioning the support of multiple radio access technologies (including 3G/ LTE/ Wi-Fi) to provide universal coverage and a seamless user experience for all 5G use cases as illustrated in Figure \ref{Fig_Fig_Sce}. 


\subsection{5G GAIA Architectural Building Blocks}

There is a need for the upcoming 5G architecture to be innovation friendly. 5G researchers can leverage the various well-known Internet engineering principles (such as indirection, modularity, resource pooling, decoupling, and extensibility, which have served as the guiding principles for Internet's architectural design) for formulating of the right 5G architecture \cite{Ghodsi2011}. In the following we discuss the key 5G architectural innovations toward global Internet access.

\subsubsection{User-Centric 5G Design}
\begin{figure}
\centering
\includegraphics[width=.4\textwidth]{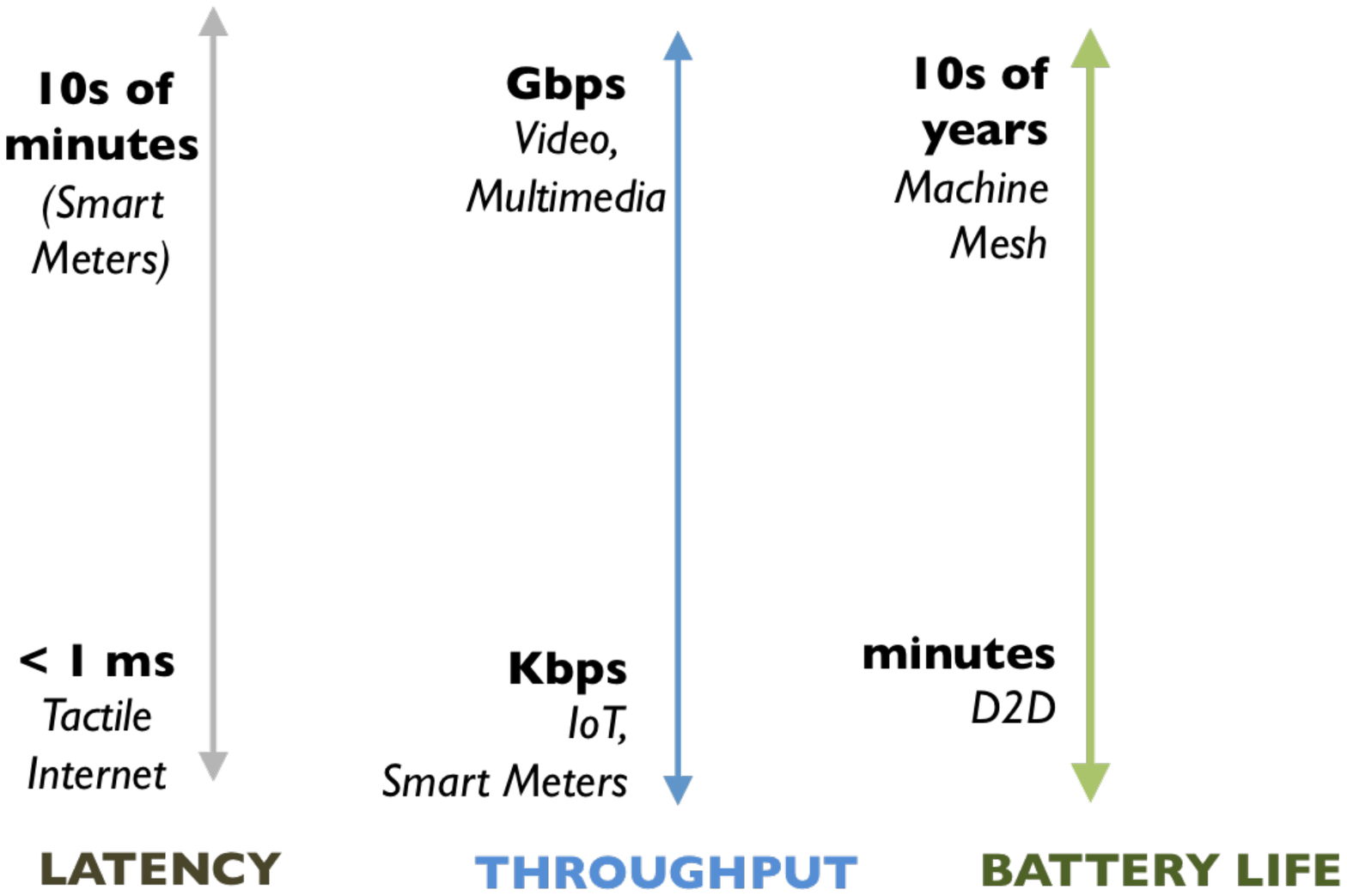}
\caption{5G can exploit the great diversity in application requirements and employ context-appropriate tradeoffs for universal provisioning.}
\label{Fig_Fig}
\end{figure}

The requirements of applications and users of 5G vary greatly (e.g., see Figure \ref{Fig_Fig}). As a result, a user-centric 5G design is very essential for universal service since the applications' requirements (latency, throughput, battery life) have significant impact on the type of deployment implemented and effectively the cost (CAPEX and OPEX) associated with such deployment. In recent times, the need of a user-centric design, instead of a cell-centric design, has led to the conceptualization of amorphous cells, decoupled signaling and data, and decoupling of uplink and downlink \cite{ChihLin2014}. Such a flexible user-centric design is amenable to energy savings (and thus economical operation) since it can facilitate flexible control of infrastructure (such as as completely turning off small cells when there is no data traffic).

\subsubsection{Green and Soft 5G}

Some key areas of research for going green and soft in 5G are identified below. The interested reader is referred to \cite{ChihLin2014} for a more elaborate treatment. 

\textit{Control-Data Plane Separation Architecture:} 5G can function more efficiently if we can enforce a control and data plane separation. The control plane functionality can be provided by a macro cell operating below the 1 GHz spectrum; alternatively, satellites and aerial platforms can also be used to provide control plane functionalities \cite{Evans2015}. The data plane, on the other hand, supports high data rate transmission and is composed of the small BSs \cite{Mohamed2016}. Since the control and data planes are separated and are not necessarily handled by the same node, the small cells (in rural areas) can be activated by the control plane on demand (or based on known user traffic profile) to deliver User Equipment (UE)-specific data when and where needed. The on-demand activation leads to significant energy savings and interference reduction. Moreover, the control-data plane separations in the rural region can be supported by the cost per connection business model, where a much lower connection cost is associated with the low rate control plane and a high connection cost for the data cell connection. 

\textit{Rethinking Signaling and Control for Diverse Traffic:} There is a need to rethink signaling and control to support all kinds of traffic on 5G (and not only voice and video). For example, a number of bursty short-message traffic types (e.g., instant messaging) have emerged that transit quickly between connected and idle states, and thus lead to increased battery drainage. It has been shown in literature that different applications have widely different ``data to signaling/control ratio'' (DSR). Current mobile networks support a single inflexible kind of signaling. For supporting universal traffic, we need to adopt various tradeoffs, and to facilitate such a tradeoff, multiple signaling styles may be needed. 

\textit{Rethinking Network-Centric Clustering and Resource Allocation:} The traditional static clustering approach is based on a fixed topology and is unable to deliver much gain for the rural networks where majority of the population lives in clusters (hamlets and villages). The rural networks will have user deployed small cells and dynamically changing network topology due their on demand activation of small cells, spatio-temporal distribution of users and service demands. Dynamic user-centric solutions, which exploits the input from self-organizing network platforms for optimal coordinated multi-point (CoMP) clustering, can be utilized for maximizing performance metrics such as energy efficiency, spectral efficiency, load balancing and user fairness, and thereby improve rural coverage. 

\textit{Self Organizing Networks (SON):} The performance of the 5G can be enhanced by employing self-organizing network (SON) features for substantial reduction in CAPEX \& OPEX by replacing the manual operation process that have been executed in legacy cellular networks with autonomous SON functions such as as self-configuration, self-optimization and self-healing \cite{Onireti2015}.

\textit{Cloud-RAN (Centralization \& Resource Pooling):} In Cloud-radio access network (C-RAN), the baseband units (BBU) of multiple BSs are migrated to a datacenter hosting high performance DSP processors. The adoption of C-RAN architecture can utilize resource pooling of substantial energy savings (e.g., saving of 70\% in OPEX of base station (BS) infrastructure has been reported in practical trials) \cite{Checko2015}, which is essential for delivering universal Internet access.

\subsubsection{Small Cells} \label{SCTech}

According to conventional understanding, rural coverage suffers in business case terms due to the low (mean) population density that is available to fund cell site investment. However, closer inspection of the rural population distribution show that most rural population lives in clusters---hamlets, villages and towns---where the local population is actually at suburban level. Hence, cells which can provide coverage and capacity at a cost that scales down quicker than the traditional solutions are the preferred in rural areas. Repurposing small cell solutions such as metrocell for rural applications (\textit{meadowcells} \cite{Brown2015}) allow the total operational cost to be scaled down by a factor of one tenth of the conventional macrocell or even lower. Nonetheless, the spread of a few small cells over a large area presents a critical problem of finding the right backhaul approach to connect the small cells to the operator's core network. The right backhaul solutions that satisfy the application requirement should be selected. The cost of such backhaul can be shared between the rural community and the operator as proposed in TUCAN3G \cite{TUCAN3G}.

\subsubsection{Do-It-Yourself (DIY) Networking}

The traditional mobile carrier deployment model is not well suited to rural areas due to the huge drop in potential revenue per square mile. The estimated revenue can nosedive by some three orders of magnitude: dropping from approximately \$250,000 per square mile of service for major urban centers to as low as \$250 per square mile for the least densely populated areas \cite{erikssonanyone}. Community networking offers a way of organically growing a community-driven network \cite{SIMO2015} that is economically attractive and provides feasible incentives to all stakeholders. With the recent emergence of low-cost software defined radios (SDRs), along with the availability of open-source software such as OpenBTS, it has now become feasible to develop low-cost community cellular networks in rural areas that have traditionally been ignored by mobile telecom companies \cite{K.Heimerl2013}. 

\subsubsection{Low Cost Spectrum Utilization}

\textit{White Space}: TV white space (TVWS) and GSM white space (GSMWS) are well suited for providing rural broadband Internet access. This is because their spectrum space is free, i.e., there is no cost associated with using their spectrum, as the user has no long-term rights to it. By using a spectrum database manager that has accurate data, the white space radio can be identified and used to provide broadband Internet access, while operating harmoniously with the surrounding channels. Contrary to the traditional Wi-Fi router, which is limited in range due to the Wi-Fi spectrum, the TVWS and GSMWS spectrum can cover 10 km in diameter, passing through blockages/obstacles such as trees, rough terrain and buildings. This range can be further extended by the use of high-gain directional antennas. With the higher range, fewer number of towers will be required to provide rural coverage, and hence, a further reduction in cost. 


\textit{Visible Light Communications (VLC)}: The new requirements of amorphous cells, separated uplink and downlink, and decoupled control and data places will require and benefit from innovations (such as user-centric visible light communication (VLC) design). VLC design is particularly interesting since LEDs will perceivably dominate the illumination/ lighting market and the piggybacking of data on LEDS modulation through at a frequency higher than the human eye's fusion rate can simultaneously facilitate the dual goals of illumination and communication. VLC uses unlicensed spectrum, has vast bandwidth, and can benefit from a ubiquitously available lighting infrastructure \cite{Zhang2015}. 

\subsubsection{Location and Time Decoupling}

\textit{Information Centric Networking (ICN):} Contrary to the traditional host-centric approach, where access to content is mapped to its fixed location, ICN eliminates this mapping and support access irrespective of the location where the content is held. The decoupling of the content and location removes the current end-to-end client server model as services and content are served by nodes that can offer such at a given point in time. This enables efficient resource allocation, as the networks storage capacity and computational resources can be jointly optimized. In addition, ICN also creates the much-needed flexibility for cost effective universal Internet access. 
\newline\textit{Delay Tolerant Device Transport:}
For delay tolerant Internet services, such as FTP and peer-to-peer file transfer, email access, over-the-air software updates, status updates from social sites and RSS feeds, mobile terminals can postpone the transmission of information messages while in transit and only engage in communication at locations with the cell with good channel condition \cite{Kolios2014}. Such store and carry forward approach can be used by Internet application, in the rural region where the parcel delivery vehicle can serve as the relay node distributing content.

\subsubsection{Smarter Devices and Edge Computing}


\textit{D2D communication}: With D2D communication, devices can directly communicate with each other without routing the data paths through a network infrastructure \cite{Serval}. D2D offers many potential gains \cite{boccardi2014five}, including: \textit{capacity gain} from sharing the spectrum resources from the cellular operator; \textit{user data rate gain} and \textit{latency reduction} due to the close proximity of the users. Though its uses cases are limited to proximity based services and the availability of content in the participating devices, it is very economical as it is exempt of the network access cost/charges which can be on the high side in rural areas.
\newline
\textit{Local Caching of data:} Intelligent data caching can be performed in a network by monitoring the data request trends of the network nodes. With the local caching of data, the bandwidth cost of Internet access is significantly reduced. Caching can be at the device (used for D2D communication), edge network (small cells), radio access network (RAN), and evolved packet core (EPC) levels. 
\newline\textit{Edge Computing:} Edge computing is a distributed computing infrastructure which requires that some application services are processed at the edge of the network in a smart device while other application services are processed in the cloud (remote data center). This leads to an increase in efficiency and a decrease in the amount of data that needs to be transported for processing, analysis, and storage. Hence, similar to local caching of data, edge computing results in a reduction in bandwidth cost of Internet access. Significant improvement in the performance of rural networks can therefore be achieved by having the much-needed application services processed at the network's edge.

\subsection{Access/ Backhaul Solutions for 5G GAIA} 

There is a whole gamut of technologies that can be used to provide access and backhauling service for 5G (with unique pros and cons). Broadly speaking, there are two well-established techniques for providing wireless communication: terrestrial-based systems and aerial/satellite-based systems. 


\subsubsection{Terrestrial Multi-hop Network}

With terrestrial networking infrastructure, the devices get good latency, but the transmitted signals suffer from scattering and multipath effects that limits the communication capacity. In such settings, a cellular coverage structure is adopted and coverage is deliberately restricted to allow the reuse of radio frequency---this however has the undesirable consequence of requiring too many antenna towers and base stations (which then have to be connected through wired or microwave links). The terrestrial networking model is now coming to a saturation point where due to a number of logistics/ costs issue it is no longer feasible to build more and more macrocells. While terrestrial networks have served us well generally, they suffer from problems. In particular, their use in rural areas is not economical. With the high OPEX and CAPEX associated with aerial platforms and satellite, the use of terrestrial multi-hop wireless networks for providing backhaul solutions has received significant attention. In the terrestrial multi-hop network, unlicensed low cost wireless network technologies (such as Wi-Fi and WiMAX) are utilized for long distance backhaul link to rural 5G small cells. Hence, it has become a very popular approach for bridging the rural-urban digital divide.

\textit{Multi-hop Wi-Fi over long distances (WiLD) network:} 
In this approach, Wi-Fi-based long distances (WiLD) links are utilized to extend Internet connectivity from the gateway node to the under-served regions and remote locations via a few number of hops. WiLD links are point-to-point wireless links with line of sight (LOS) over long distances such as 10-100 km and high-gain directional antennas. The use of Wi-Fi for providing coverage for the under-served regions and remote locations is triggered largely by the extensive availability of IEEE 802.11 hardware at very low cost and low power, and the fact that it is operated license-free at the industrial, scientific, and medical radio band (ISM band). In addition, the nodes are light in weight and therefore do not require expensive towers. 

\textit{Multi-hop network based on WiMAX links:}
WiMAX standards though conceived for fixed metropolitan area networks (MAN), also offers solutions that are promising for the long-distance communication in rural broadband networks. Similar to WiLD, WiMAX also requires high towers in order to ensure line of sight for the long-distance links in flat rural areas. Having LOS is of utmost importance due to the restriction in the transmission power in non-licensed bands. 

\subsubsection{Satellites and Aerial Platforms}
 
In rural settings, satellites can provide much broader wireless coverage with considerably lesser terrestrial infrastructure. Satellites---which exist in many varieties such as (1) geostationary (GEO) satellites; (2) medium-earth orbit (MEO) satellites; and (3) low-earth orbit (LEO) satellites---are not without problems too. GEO satellites typically require expensive and bulky user equipment. LEO and MEO satellites, on the other hand, increase system complexity through the fact that they are not static with respect to the user. In recent times, various configurations of systems that deploy aerial networking infrastructure (such as small UAV drones, balloons, low-altitude platforms (LAPs) and higher-altitude platforms (HAPs) have been proposed---each of which provides a unique bandwidth/ latency/ price/ performance tradeoff. These aerial networking technologies can not only augment existing infrastructures but can also provide wireless service where currently none exists. Companies such as Facebook, Google, O3B are currently engaged in rolling out projects that utilize aerial networking systems for the purpose of universal Internet accessibility and GAIA. Just like drones have made a marked impact on the fields of logistics, defense, and safety, the use of drones---and other aerial networking systems---promise to disrupt wireless networking, leading the way to improved network performance and better flexibility in emergency situations. Some important issues to utilizing and integrating satellites and aerial platforms for 5G-based GAIA are discussed next. 

\textit{Coverage:} In line with the ubiquitous coverage aim of 5G networks, satellites and aerial platforms can provide wide coverage to either complement or extend the dense terrestrial cells. Though they cannot match the area spectral efficiency of the terrestrial 5G networks, they can provide larger cells which can serve as the control plane in a control-data separated architecture and thereby relieve the terrestrial network of management and signaling functions. Furthermore, their larger cells can be used in critical and emergency services.

\textit{Backhaul:} As earlier identified in Section \ref{SCTech}, backhaul is a major issue with the geographically spread small cells in rural areas. High throughput satellites and aerial platforms can be used to provide backhaul in areas where it is difficult or cost ineffective to do so terrestrially. In addition, in a virtualized network, some of the network node functions can be included on-board the aerial platforms or satellites and thereby save in the cost of the on-ground sites \cite{Evans2015}.

\textit{Integration:} The integration of aerial platforms and satellites with terrestrial systems can lead to great gains in 5G. The quality of experience can be improved by intelligent traffic routing and data caching for onward terrestrial transmission \cite{Evans2015}. Also, as a result of intelligent caching, the propagation latency with satellites and aerial platforms will no longer be an issue.

\subsubsection{Access/ Backhaul Challenges} 
The main challenges of access and backhaul networks in achieving universal Internet goal are as follows:

\textit{Cost}: The backhaul cost, particularly with small cells, is a significant portion of the overall cell cost. The backhaul solution should be optimized while considering the number of backhaul nodes, and subject to minimizing the CAPEX and satisfying the UE's Quality of Experience (QoE). In scenarios where shared backhaul links are used to serve several small cells or where multiple backhaul solutions are available, self-organizing backhaul algorithms can be utilized to optimize and automate the backhaul configurations of the small cells as they are deployed \cite{Jaber2015}. 


\textit{Sychronization:} Time and frequency synchronization are required to ensure that the transmitted data utilize their allocated channel and comply with the system specifications. In outdoor environment, global navigation satellite system (GNSS) can be used to provide accurate time and frequency synchronization. However, this approach may not work in indoor environment or outdoor environment with limited sky view. 


\textit{Capacity and Energy:}
The capacity of the backhaul solution must not constrain the small cell capacity and should also have a good margin that can provision for statistical variation and future growth \cite{Jafari2015}. Also, most of the developing nations have the energy supply problem, which is a major challenge in powering up the access and backhaul networks. Renewable solutions such as solar and wind energy are alternatives.

\textit{Scale:}
The rural network scale (area and number of devices) tends to lead to a lower Average Revenue Per User (ARPU) yield compared to the urban market. This makes the investment in rural access and backhaul connectivity doubly challenging, and as a result, operators find it quite difficult to build a business case to roll out their own backhaul infrastructure. 


\subsection{Pre-5G Cellular Projects for Rural Areas}

There has been limited work on developing technically and economically feasible broadband solutions for rural areas in which the users have conventional 3G/4G cellular terminals. The EU-funded TUCAN3G project proposed to utilize 3G femtocells (called ``Home Node B'', or HNB in 3G terminology) in outdoor environments and to use heterogenous backhauling using technology such as Wi-Fi for long distance (WiLD) and WiMAX in unlicensed bands \cite{TUCAN3G}. In TUCAN3G, the initial investment cost of providing the 3G services in the rural community is reduced by sharing the community-owned backhaul with the operator. This sharing approach is beneficial for both the operator and the community, since the latter benefits by using the revenue generated to maintain its network while also enjoying from having 3G services. On the other hand, the operator gains by avoiding CAPEX, thus making it affordable for them to offer 3G services to rural areas where the revenue per square mile is quite low.

\section{Discussion}

\subsection{Cellular Architecture and GAIA}

Since cellular technologies have traditionally not prioritized coverage as a primary goal, many of the adopted design choices are not perfectly in sync with the needs of GAIA. We can even question if the cellular architecture itself---which has has survived every new generation of mobile standards till 4G in one form or the other---is suitable for GAIA. Despite its longevity, the cellular architecture has never been successful in cost-effectively bringing service to the low-average revenue per user (ARPU)/rural areas. Furthermore, the implicit assumption of homogeneity made by cell-centric design has meant that various short fixes---such as relays, CoMP, distributed antenna systems (DASs), and heterogeneous networks (HetNets)---had to be invented as short-term solutions. With 5G now in a formative phase, the time is right to critically evaluate the suitability of cellular architecture for high performance, flexibility, universal access, and economic operation. 

\subsection{Major 5G Technologies and GAIA}

We can also evaluate other 5G approaches for their alignment with GAIA. The three principal approaches being explored by 5G researchers are: 1) \textit{ultra densification}, in which we provision more cells in a given area using technologies such as pico and femtocells; 2) \textit{mmWave}, in which we leverage more bandwidth by moving to the mmWave spectrum and; 3) \textit{massive MIMO}, in which we increase spectral efficiency via large-scale antenna systems \cite{Andrews2014}. All of these three techniques are primed towards increasing the peak data rate but are found wanting when we evaluate them for the purpose of GAIA. In terms of ultra-densification, cost and interference has been identified as its major challenge in urban deployment \cite{Monserrat2015}. The cost issue becomes even more overbearing in rural/low-ARPU regions, thus making the use of such ultra-dense 5G solutions ill suited for GAIA. The use of the mmWave band similarly is problematic for GAIA due to the higher path loss at the mmWave band that effectively reduces the coverage to smaller areas. The use of mmWave band resultantly implies an increase in CAPEX (either due to the requirement of more mmWave base stations to achieve coverage, or due to the need of implementing high-complexity beamforming to overcome the attenuation) \cite{Monserrat2015}. Finally, even though massive MIMO can provide great gains in rural/low-ARPU settings, the computational burden associated with it makes it unaffordable to the rural/low-ARPU region.

In the light of the observations above, we posit that coverage and universal access should also be considered as important performance metrics that we use to judge technologies for 5G. 


\subsection{Optimizing Protocols for 5G Internet}

Considering a future world in which 5G will be used to drive GAIA forces us to prioritize research on optimizing the Internet experience for users that connect to the Internet through mobile connections \cite{gitau2010after}. A lot of research has shown that some characteristics of cellular networking differs drastically from those of typical data networks \cite{shafiq2011characterizing}. For instance, cellular networks can suffer from `bufferbloat'' that can result in several seconds of round trip delay and other performance degradation \cite{Jiang2012}. In order to alleviate this problem, cellular-focused cross-layer congestion protocols such as CQIC, which attains a low round trip delay values across a range of flow sizes can be adapted \cite{Lu2015}. CQIC achieves its gains by using the discontinuous transmission ratio and channel quality indicator to predict the instantaneous cellular bandwidth.


\subsection{New 5G Performance Metrics}

There is also a need of developing new \textit{coverage-incorporating} performance metrics for 5G. It is worth highlighting that universal coverage is not only useful for rural users but also is valuable for ensuring high quality of experience for business users (who often travel extensively and require high-quality services during their mobility). 

\subsection{Adopting Context-Appropriate Tradeoffs} 

We envision that 5G technology---if it has to scale to global proportions and be affordable to all---will require unprecedented flexibility and evolvability of its network architecture to cope with the myriad technologies and the broad spectrum of application requirements (see Figure \ref{Fig_Fig}). Simply speaking, there is \textit{no one-size-fits-all networking solution} that will work to provide high-quality affordable service for all of humanity. This makes the use of tradeoffs inevitable. To manage the tradeoffs appropriately according to the context, we propose the use of the ``\textit{approximate networking}'' approach \cite{qadir2016an} through which \textit{context-appropriate tradeoffs} are adopted to pragmatically provision universal Internet services with 5G. The quality metrics considered in networking include: latency, throughput, fidelity, convenience, performance, cost efficiency, coverage, reliability, privacy, free content/services.  Context-appropriate tradeoffs can consider any combination of the metrics in order to achieve universal Internet access. \cite{Onireti2011,Capps2009,Vulimiri2013}.

Acheiving the GAIA goal requires flexibility in defining and implementing the ambitious 5G data rate goals. While an ideal 5G Internet experience entails the simultaneous optimization of multiple metrics (minimizing congestion, delays, errors, along with the maximization of reliability, bandwidth, and capacity), the practical realization of such an ideal 5G service will be costly (and thus out of the reach of many potential subscribers worldwide). In this regard, 5G designers should consider the following two perspectives to capacity provisioning \cite{Brown2015}. Firstly, the rate should be set according to what is economically achievable; this entails presenting the same system capacity with various level of speed, which could range from 5 Mb/s to 50 Mb/s, and associating various percentage of time and coverage probability to locations. Secondly, a perception of infinite capacity can be created for the user by delivering an ``always sufficient'' response to instantaneous demand; this requires the careful management of the quality of experience (QoE) and the available resources. 


\subsection{5G, Energy Efficiency, and GAIA} 

While the bulk of research attention has focused on increasing spectral efficiency (SE) for higher capacity, the focus on energy efficiency (EE) has received less attention---although recent works have started emphasizing joint EE and SE optimization \cite{Onireti2013,Onireti2011}. EE optimization is especially important for universal service since economics will play a large role when we consider developing/ rural regions. The need of optimizing for EE will become ever more important when we consider the fact that by over-exhausting the vital \textit{non-renewable} natural resources (such as fossil fuels), our society's cheap fossil fuel fiesta will likely end soon, leading the bulk of humanity towards the long arc of scarcity, constraints, and limits as soon as we cross the tipping point of global oil production \cite{qadir2016taming}. This state of \textit{undevelopment} is likely to affect both developed countries and developing countries. This brings to the fore the importance of designing energy efficient 5G solutions that are  optimized for scarcity and constraints. We envision that in such constraints-afflicted environment, throughput will cease to be the all important 5G performance metric, and the concept of context-appropriate tradeoffs and approximate networking will become more important \cite{qadir2016taming}.

\section{Conclusions}
In this paper, we have presented an overview of the following 5G solution for universal Internet access: green and soft 5G, control-data plane separation architecture, smarter networks, small cell technology, bottom-up DIY community networking and low cost spectrum utilization. We highlighted the 5G-based GAIA access/backhaul issues and challenges. We have identified context-appropriate trade-off as an approach to achieve universal Internet access in 5G.

{\scriptsize
\bibliographystyle{abbrv}
\bibliography{sigproc}
}

\end{document}